\documentclass[preprint,showpacs,preprintnumbers,amsmath,amssymb,prl]{revtex4}

\usepackage{graphicx}
\usepackage{dcolumn}
\usepackage{bm}
\usepackage{natbib}
\usepackage{amssymb}
\usepackage{epsfig}
\usepackage{graphics}
\usepackage{graphicx}

\begin{document}

\title{Lanthanide impurities in wide bandgap semiconductors:
a possible roadmap for spintronic devices}

\author{G. Caroena$^{\rm (1)}$, W. V. M. Machado$^{\rm (1)}$, 
J. F. Justo$^{\rm (2)}$, and L. V. C. Assali$^{\rm (1)}$}

\affiliation{$^{\rm (1)}$ Instituto de F\'{\i}sica,
Universidade de S\~ao Paulo,\\
CP 66318, CEP 05315-970, S\~ao Paulo, SP, Brazil \\
$^{\rm (2)}$ Escola Polit\'ecnica, Universidade de S\~ao Paulo,\\
CP 61548, CEP 05424-970, S\~ao Paulo, SP, Brazil}

\begin{abstract}
The electronic properties of lanthanide (from Eu to Tm) impurities in 
wurtzite gallium nitride and zinc oxide were investigated by first 
principles calculations, using an all electron methodology plus a 
Hubbard potential correction. The results indicated that the 4f-related 
energy levels remain outside the bandgap in both materials, in good 
agreement with a recent phenomenological model, based on experimental 
data. Additionally, zinc oxide doped with lanthanide impurities 
became an n-type material, showing a coupling between the 4f-related 
spin polarized states and the carriers. This coupling may generate spin 
polarized currents, which could lead to applications in spintronic devices. 
\end{abstract}


\maketitle

Wide bandgap semiconductors, doped with rare earth (RE) impurities, received 
great attention over the last decade, mainly due to potential applications in 
optoelectronic \cite{agazzi} and spintronic \cite{hite,davies} devices. 
Rare earth lanthanide atoms exhibit partially filled 4f shells, 
and intra-4f electric dipole transitions are forbidden for the free ions. 
On the other hand, the respective transition probabilities increase 
considerably when the ions are placed in a crystal field, that splits 
the 4f-related energy levels, modifying the dipole selection rules and 
leading to luminescent centers \cite{steckl}.

Although theoretical modeling has been used to investigate the
properties of RE-related materials \cite{petit,sanna}, several 
methodological challenges have hindered an appropriate description of 
the 4f-related states in crystalline environments, such as RE 
impurities in semiconductors. It is well documented 
that the density functional theory generally fails in describing 
highly correlated systems, such as the interactions in 4f-related  
electronic states. Such limitations could in part be overcome with 
the introduction of an on-site Hubbard U potential 
correction \cite{anisimov,novak,liech}. One of the major outcomes of 
this correction is a better description of the energy splitting 
between occupied and unoccupied 4f-related electronic levels. 
We have recently shown that 
this procedure provides a good description of the electronic structure
of metallic RE crystals \cite{drm}, when  compared to available 
experimental data \cite{lang}. Here, we show that the same procedure
is also appropriate to describe the trends on the 4f-related energy levels of RE 
impurities (from Eu to Tm), with respect to the bandgap, in 
wurtzite gallium 
nitride and zinc oxide crystals. The results were discussed in the 
context of a recent phenomenological model to determine the energy 
positions of 4f-related electronic systems in crystalline 
environments \cite{dorenbos12}. Our results also indicate that
doping ZnO with lanthanide impurities leads to a n-type material, 
with a magnetic coupling between the 4f-related states and the 
carriers. This suggests that such systems could generate spin 
polarized carrier currents, allowing to envision potential 
applications in spintronic devices.

In RE lanthanide atoms, the 4f states play a minor role on bonding, 
staying in an atomic-like configuration. Therefore, those atoms bind to 
their neighboring ones through their outer (5d, 6s, and 6p) atomic 
valence states \cite{lang}. Earlier theoretical investigations 
generally considered the 4f electrons as core states, with constrained 
occupations \cite{pet} which were based on experimental results. State of the 
art methodologies, such as the one used here \cite{singh,blaha}, now 
allow to treat the 4f electrons as valence states, with no constrained 
occupations, opening the possibility to model those systems more 
realistically.

Our calculations were performed using the all-electron spin-polarized 
full-potential linearized augmented plane wave method \cite{singh}, 
implemented in the WIEN2k package \cite{blaha}, within the framework 
of the density functional theory. The exchange correlation potential
was described by the generalized gradient approximation \cite{pbe} 
plus the on-site Hubbard U potential, implemented in a rotationally 
invariant procedure \cite{anisimov,novak,liech}. Self-consistent 
iterations were performed until convergence in the total energy of 
10$^{-5}$ Ry was achieved. In all systems, the internal degrees of 
freedom were optimized, with no symmetry constraints, until the force 
in any atom was smaller than 10$^{-3}$ Ry/a.u. We considered a 56-atom 
reference hexagonal supercell of GaN or ZnO, with the RE atom in a 
substitutional cation site. A detailed description of the simulation cell
is presented elsewhere \cite{assali2004}. In order to check the
convergence of our results with respect to the supercell size, we
performed test simulations with larger supercells (up to 108 atoms)
and observed that the results were essentially unchanged in comparison
to those with the 56-atom supercell. The irreducible Brillouin zone was 
sampled by a ($2 \times 2 \times 2$) grid. Convergence on the total 
energy was achieved using a plane-wave basis set, limited by the wave 
number 8.0/R (maximum length of the plane-waves), where R = 1.5 a.u. 
is the smallest atomic sphere radius. All those approximations provide 
a reliable description on the electronic and structural properties of 
several impurity centers in semiconductors \cite{assali2004,bn,larico,assali}.

We initially used an on-site Hubbard potential correction, computed
self-consistently \cite{novak}, for the 3d states of the cation atoms (Ga or Zn, 
respectively in GaN or ZnO) in the crystal hosts. We found self-consistent 
values of $\rm U_{3d} (Ga) = 10.9$ eV and $\rm U_{3d} (Zn) = 7.6$ eV. 
Such corrections provided structural 
and electronic properties of the respective crystalline systems in 
agreement with available experimental \cite{landolt,magnuson10} 
and theoretical \cite{janotti} data. The Hubbard correction improved
the description of the bandgap of GaN (ZnO), going
from  1.7 (0.8) eV to 2.2 (1.6) eV. This correction still underestimates
the bandgap of GaN (ZnO) as compared to the experimental value of 
3.5 (3.4) eV \cite{landolt,magnuson10}.  We then implemented the 
correction in the 4f states of the RE impurities within the
same methodology, in order to get an appropriate description of the 
highly correlated 4f-related electronic states.

\begin{figure}[h!]
\centering{
\includegraphics[width=12.0cm]{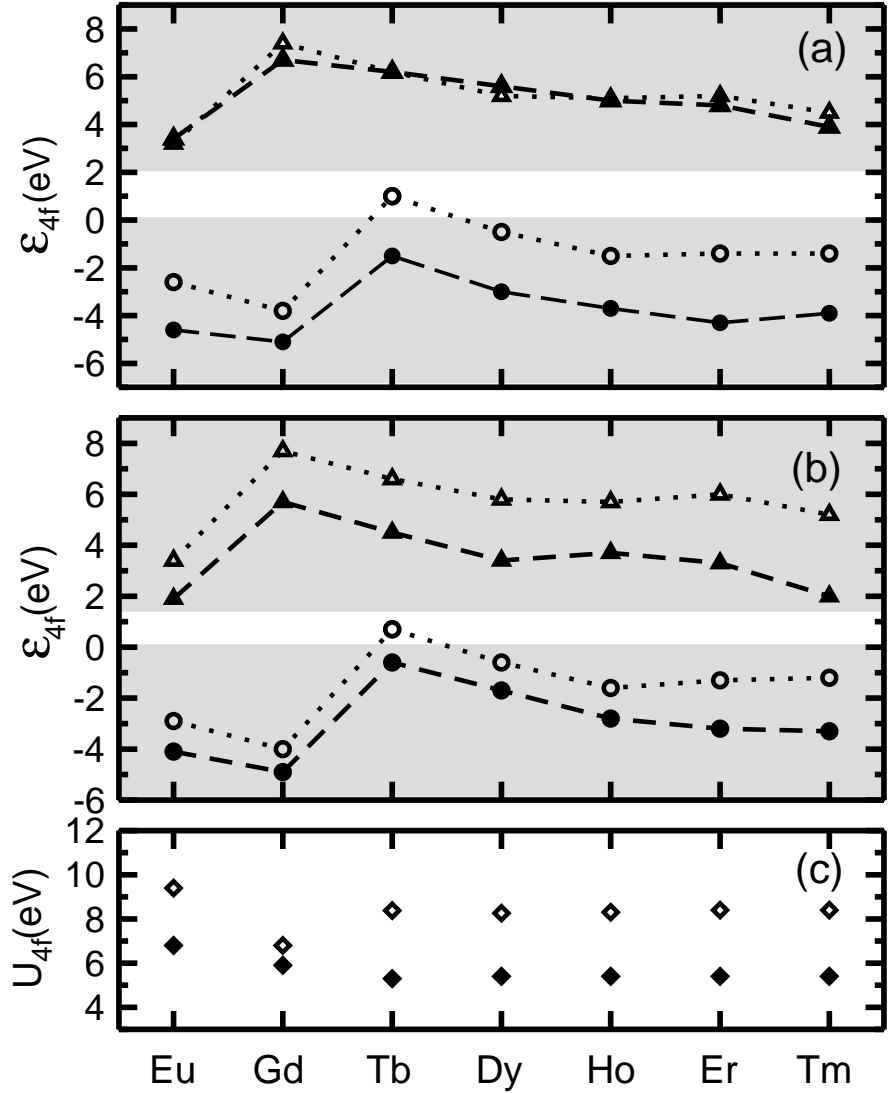}
\vspace*{1cm}
\caption{Results of our calculations (closed symbols) of RE impurities
in (a) GaN and (b) ZnO compared to the ones predicted from a 
phenomenological model (open symbols) \cite{dorenbos06,dorenbos07}. The figure
presents the highest occupied (circles) and 
lowest unoccupied (triangles) 4f-related energy eigenvalues 
($\varepsilon_{\rm 4f}$), in the respective valence and conduction bands
(gray regions), considering the Hubbard potential corrections. 
All experimental and theoretical values are presented with respect to
the GaN and ZnO valence band tops, taken as reference values.
The (c) panel presents the $\rm U_{4f}$ values,  obtained self-consistently,
for 4f-related states of each impurity in GaN (open diamond symbols) and ZnO 
(closed diamond symbols).}
\label{fig1}
}
\end{figure}

Figures \ref{fig1}(a) and (b) present the highest occupied (HO) and lowest 
unoccupied (LU) 4f-related RE eigenvalues with relation to the GaN 
and ZnO bandgaps, respectively. For GaN, shown in fig. \ref{fig1}(a), 
our results give trends along the RE series in excellent 
agreement with those of a phenomenological model based on experimental 
data \cite{dorenbos03,dorenbos06}. However, our calculations 
provide absolute values for the HO levels which are about 2 eV
lower when compared to that model. Our results also indicated that RE 
impurities introduce no energy levels within the materials bandgap, 
which is fully consistent with the phenomenological model, except for 
the Tb impurity. In order to clarify this point, we stress that a direct 
comparison of our results with experimental or phenomenological ones
has its shortcomings. 
Our results are represented by the 4f-related Kohn-Sham eigenvalues, 
while the experimental data comprises the multiplet electronic 
configurations, measured for the electronic transitions, related
to the respective ionization energies. However, our 
results provided a proper description on the 4f-related energy splitting 
(excitation) between the 3+ and 2+ oxidation charge states of all RE
impurities, which 
is represented by the respective energy difference between the HO
and the LU 4f-related states 
($\rm \Delta \varepsilon_{4f}$). These 
results offered an additional certification that the procedure used here, 
to obtain the Hubbard U values for the 4f-states, describes their 
electronic correlation appropriately.

Figure \ref{fig1}(c) presents the Hubbard $\rm U_{4f}$ values,
obtained self-consistently, for each impurity in GaN and ZnO. For GaN, those 
values are very close to the respective $\rm \Delta \varepsilon_{4f}$, 
all of them lying between 8 and 9 eV. The only exception is Gd, where 
$\rm U_{4f} (Gd) = 6.8~ eV$ and $\rm \Delta \varepsilon_{4f} (Gd) = 11.7~ eV$.
The Gd impurity is a particular case in the RE series, since it has a 
half-filled 4f state. As a result, $\rm \Delta \varepsilon_{4f}$  
is between a fully occupied spin up energy level and a fully
unoccupied spin down one. For any other impurity of this series, 
$\rm \Delta \varepsilon_{4f}$ is between two spin up levels (before Gd) 
or two spin down ones (after Gd). Therefore, the exchange-correlation
interaction for Gd, and consequently the energy splitting between occupied
and unoccupied states, is already reasonably well described even without 
a Hubbard correction. Without the correction, $\rm \Delta \varepsilon_{4f} (Gd)$
is already 5.2 eV in our calculations, fully consistent with a 
recent investigation \cite{dalpian}, that found a value around 4.5 eV. 
For any other RE impurity, without the Hubbard correction, 
$\rm \Delta \varepsilon_{4f}$ is smaller than 1 eV. As a result, the 
electronic structure of those systems cannot be described appropriately 
without the Hubbard correction.  Therefore, a smaller $\rm U_{4f}$ value 
for Gd, as compared to other RE impurities, is enough to provide a
proper description of the respective electronic structure.

Consistent with the discussion for impurities in GaN, the trends along 
the RE series in ZnO are in excellent agreement with the  phenomenological 
model \cite{dorenbos07}, as shown in  fig. \ref{fig1}(b). 
Our results indicate that no RE impurity introduces 
energy levels within the ZnO bandgap. However, the phenomenological model 
suggests that the 3+ oxidation state of Tb is inside the ZnO bandgap 
(about 0.8 eV higher than the top of the valence band). 
The trends on the splitting between the HO and LU 4f-related RE 
levels are in good agreement with that model \cite{dorenbos07}. 
As discussed in the previous paragraph for
Gd in GaN, Gd in ZnO also has an U value that is considerably smaller
than the respective HO-LU energy splitting, as shown in fig. \ref{fig1}(c). 
In ZnO, $\rm U_{4f} (Gd) = 5.9~ eV$ while 
$\rm \Delta \varepsilon_{4f} (Gd) = 10.6~ eV$.
According to the figure, the computed $\rm U_{4f}$ values 
of the respective RE
impurities are systematically smaller in ZnO than in GaN, which reflects
the role of the neighboring atoms, and even the
larger ionicity of ZnO compared to that of GaN, on the 
4f-related correlation.

It is worth mentioning that we computed the valence band offset between 
ZnO and GaN host materials, following the procedure described in Ref. \cite{bn}.
In order to get that, we considered an alignment of
the RE 5d-related energy levels, according to the average energy of spin
up and down states. For all RE impurities, this average energy is
resonant in the conduction band, at 4.7 eV (5.9 eV) above the GaN (ZnO) 
valence band top, which
resulted in a band offset of 1.2 eV between those two materials.
This band offset is essentially the same with or without the Hubbard U
correction in the 3d-states of the cation atoms. Such result is 
in excellent agreement with the one 
from another investigation \cite{walle}, providing an additional
certification of this methodology to describe the positions
of the energy eigenvalues with respect to the materials bandgaps.

The results of RE in ZnO showed great similarities with those of RE 
in GaN. Such similarities are not fortuitous and carry important 
implications. First, when a RE impurity replaces a trivalent cation atom 
(Ga) in GaN, it donates three electrons to bind with the nitrogen 
neighboring atoms, staying in a 3+ oxidation state. As a result, all the 
RE centers in the neutral charge states have their total spin associated 
exclusively to the 4f-related states. Following the same reasoning, when 
a RE impurity replaces a divalent cation atom (Zn) in ZnO, it would be 
expected that the impurity donates two electrons to bind with the oxygen 
neighboring atoms, resulting in a 2+ oxidation state. Our results showed 
otherwise, they indicated that all RE impurities in ZnO stay in a 3+ 
oxidation state, as for the RE impurities in GaN. Such results 
in ZnO are fully consistent with assumptions used in the phenomenological 
model \cite{dorenbos07}. For a RE in ZnO to achieve such oxidation state, 
it donates two electrons to stabilize the binding with the oxygen neighboring
 atoms, while the third electron populates the bottom of the ZnO conduction 
band. As a result, in contrast to what occurs in GaN, a neutrally charged
RE impurity in ZnO has its total spin associated to both the 
4f-related states and the delocalized spin polarized carrier in the 
conduction band. Therefore, doping ZnO with RE impurities leads to an n-type 
semiconductor.

\begin{figure}[h!]
\centering{
\includegraphics[width=13.0cm]{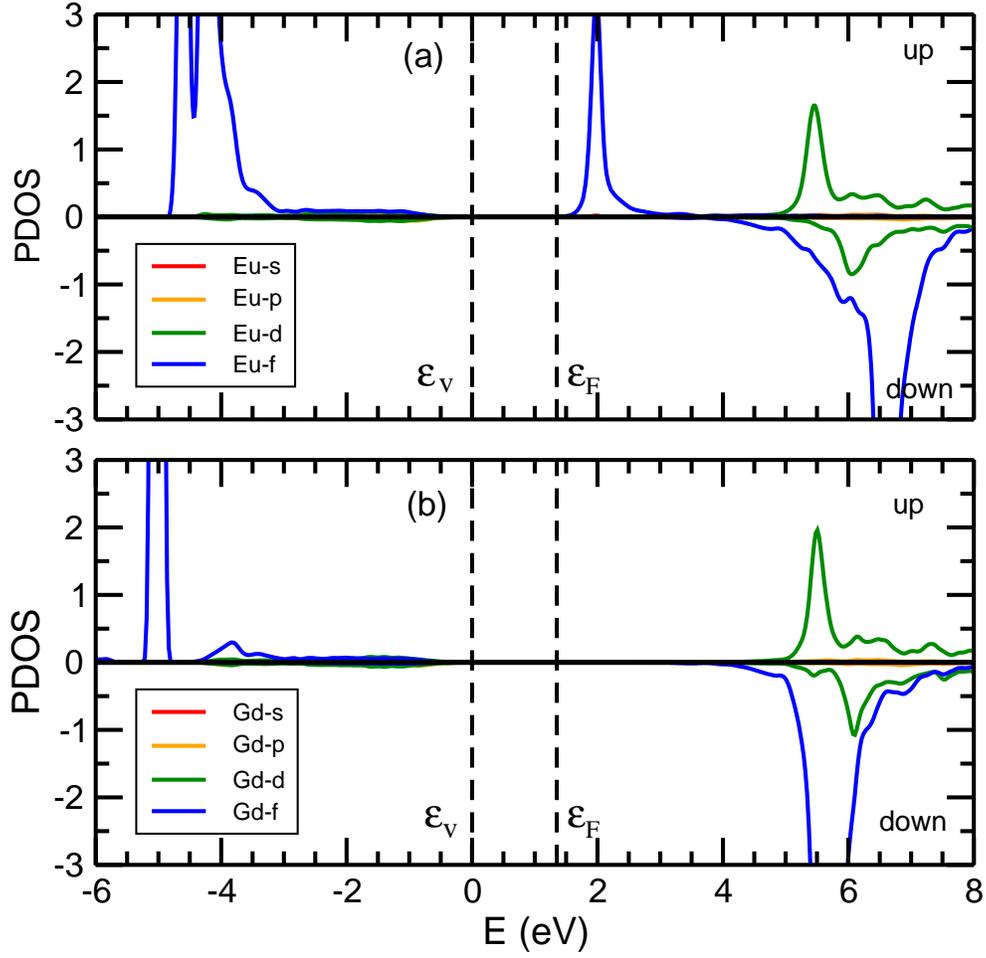}
\vspace*{1cm}
\caption{The s, p, d, and f projected density of states (PDOS)   
inside the (a) Eu and (b) Gd atomic spheres ($\rm R_{RE} = 2\ a.u.$) 
for spin up and spin down energy levels in ZnO. The dashed lines 
represent the valence band top ($\rm \varepsilon_v$)  and the 
conduction band bottom
(at the Fermi energy $\rm \varepsilon_F$).} 
\label{fig2}
}
\end{figure}

\begin{figure}[h!]
\centering{
\includegraphics[width=6.0cm]{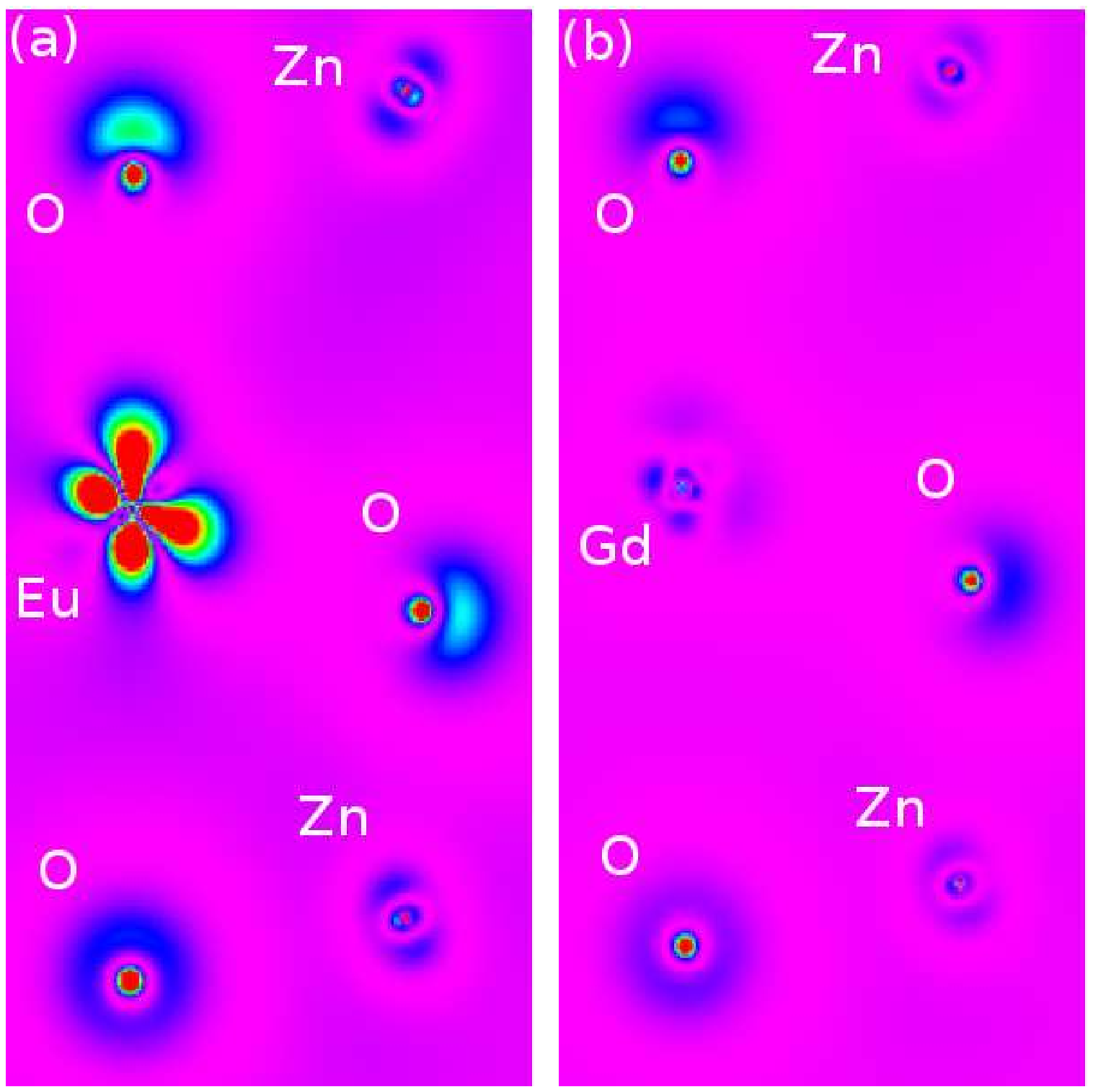}
\vspace*{1cm}
\caption{Electron density in the
$ [11 \overline{2}0]$ plane for the electron in the conduction band
bottom (Fermi energy level) for (a) Eu and (b) Gd in ZnO. The 
coloring goes from red (high density) to violet (low density), following
the rainbow sequence.} 
\label{fig3}
}
\end{figure}

The delocalized spin polarized carrier in the conduction band bottom opens 
the possibility of getting diluted magnetic semiconductors \cite{dietl} 
using RE doped ZnO. Such possibility could be investigated by observing the 
partial density of states in the impurity sites. Figure \ref{fig2} presents 
the projected s, p, d, and f density of states inside the RE atomic sphere 
of the ZnO:Eu and ZnO:Gd systems. For Eu, there is an unoccupied 4f-related 
state just over the Fermi energy, while for Gd such 4f-related state 
is about 4 eV above the Fermi energy. In the ZnO:Eu system, the unoccupied 
4f-related level interacts with the delocalized energy level that defines 
the conduction band bottom, which is occupied by an almost-free carrier. 
This result indicates the viability of getting spin polarized carriers in 
ZnO:Eu, although this would be more difficult using other RE impurities. 
Figure \ref{fig3} shows  the density of the electron (carrier) in the 
conduction band bottom for Eu and Gd impurities in ZnO. The figure shows 
the localized 4f character in the Eu site in ZnO:Eu, which is
 much stronger than in the ZnO:Gd system. For Eu, such localization
(about 7\% of charge inside the Eu sphere) is enough to provide a spin 
polarized carrier current, within the Zener model \cite{dietl}.

In summary, we have investigated the electronic properties of substitutional
RE impurities (from  Eu to Tm) in gallium nitride and zinc oxide. We found that 
the RE impurities stay in a 3+ oxidation state in both materials, consistent with 
experimental assumptions. Additionally, we have shown that the self-consistent 
procedure to compute the values of the Hubbard U parameters provides a reliable 
description on the electronic properties of those impurity centers. This is 
confirmed by comparing our results with available data from a recent
phenomenological model \cite{dorenbos12}. Our theoretical model also indicated 
that no RE impurity introduce energy levels in either GaN or ZnO bandgaps.
However, in ZnO the impurities lead to an n-type material, independent of
the Fermi level energy. There is strong 
evidence that the negative carrier could be magnetically coupled with the
4f-related states, which could generate spin polarized carrier currents. 
Moreover, since the 4f-related energy levels introduced by the impurities 
are highly localized, weakly interacting with the host atoms, the 
alloying of ZnO with these RE elements could lead to diluted magnetic 
semiconductors. For RE in GaN, our results show that the alloying of GaN  
with these elements could only lead to diluted magnetic semiconductors if 
the RE impurities are complexed with other defects, in order to generate the 
almost free carriers, as discussed in Ref. \cite{sanna}. 
\vspace{0.3cm}

\noindent
\textbf{Acknowledgments}

This work was partially supported by the Brazilian Agencies CNPq and 
FAPESP.


\vfill
\eject


\pagebreak

\pagebreak

\end{document}